\documentclass[1p]{elsarticle}
\usepackage{graphicx}
\usepackage[english]{babel}
\usepackage{multirow}
\usepackage[pdftex,svgnames,dvipsnames]{xcolor}
\usepackage{lscape}
\usepackage{pdflscape}
\usepackage{enumitem}

\usepackage{amsmath}
\usepackage{graphicx}
\usepackage[colorlinks=true, allcolors=blue]{hyperref}
\usepackage{cleveref}
\newcommand{\changed}[1]{#1}

\usepackage[pdftex,svgnames,dvipsnames]{xcolor}

\journal{Journal of Systems and Software
}

\begin{document}

\begin{frontmatter}

\title{reAnalyst: Scalable Annotation of Reverse Engineering Activities}

\author[1]{Tab (Tianyi) Zhang\fnref{fn1}}
\ead{tab.zhang@ugent.be}
\author[3]{Claire Taylor\fnref{fn1}}
\ead{claire.g.taylor.1988@gmail.com}
\author[1]{Bart Coppens}
\ead{bart.coppens@ugent.be}
\author[2]{Waleed Mebane}
\ead{mebanew@arizona.edu}
\author[2]{Christian Collberg}
\ead{collberg@cs.arizona.edu}
\author[1]{Bjorn De Sutter}
\ead{bjorn.desutter@ugent.be}

\affiliation[1] 
{
organization={Computer Systems Lab, Ghent University},
addressline={Technologiepark-Zwijnaarde 126},
postcode={9052},
city={Gent},
country={Belgium}
}
\affiliation[2]
{
organization={Department of Computer Science, The University of Arizona},
postcode={85721},
city={Tucson},
country={USA}
}
\affiliation[3]
{
organization={Lawrence Livermore National Laboratory},
postcode={94550},
city={Livermore},
country={USA}
}

\fntext[fn1]{Tab (Tianyi) Zhang and Claire Taylor share dual first authorship.}

\begin{abstract}
This paper introduces reAnalyst, a framework designed to facilitate the study of reverse engineering (RE) practices through the semi-automated annotation of RE activities across various RE tools. By integrating tool-agnostic data collection of screenshots, keystrokes, active processes, and other types of data during RE experiments with semi-automated data analysis and generation of annotations, reAnalyst aims to overcome the limitations of traditional RE studies that rely heavily on manual data collection and subjective analysis. The framework enables more efficient data analysis, which will in turn allow  researchers to explore the effectiveness of protection techniques and strategies used by reverse engineers more comprehensively and efficiently. Experimental evaluations validate the framework's capability to identify RE activities from a diverse range of screenshots with varied complexities.  Observations on past experiments with our framework as well as a survey among reverse engineers provide further evidence of the acceptability and practicality of our approach. 
\end{abstract}

\begin{keyword}
Reverse Engineering Tools, Software Protection, Man-At-The-End Attacks, Empirical Studies, Analysis tools, Image Analysis 
\end{keyword}

\end{frontmatter}

\newcounter{cccnt}
\newcounter{bdscnt}
\newcounter{tzcnt}
\newcounter{bccnt}
\newcounter{wmcnt}

\newcommand{\cc}[1]{\refstepcounter{cccnt}
	\textcolor{Sepia}{\textbf{CC [\thecccnt]:} #1}}
\newcommand{\ccB}[1]{\refstepcounter{cccnt}
	\textcolor{Sepia}{\textbf{CC[Feb13] [\thecccnt]:} #1}}
\newcommand{\bds}[1]{\refstepcounter{bdscnt}
	\textcolor{RedOrange}{\textbf{Bjorn [\thebdscnt]:} #1}}
\newcommand{\tz}[1]{\refstepcounter{tzcnt}
	\textcolor{BrickRed}{\textbf{Tab [\thetzcnt]:} #1}}
\newcommand{\bc}[1]{\refstepcounter{bccnt}
	\textcolor{Blue}{\textbf{Bart [\thebccnt]:} #1}}
\newcommand{\ct}[1]{\refstepcounter{bccnt}
	\textcolor{OliveGreen}{\textbf{Claire [\thebccnt]:} #1}}
\newcommand{\wm}[1]{\refstepcounter{wmcnt}
	\textcolor{Dandelion}{\textbf{Waleed [\thewmcnt]:} #1}}

\section{Introduction}

Understanding the practice of software reverse engineering (RE) is important for teaching it, for RE tool and strategy development, and for the development, evaluation, and validation of \changed{techniques used for} anti-RE software protections such as software obfuscation~\cite{Ceccato2014need,Dagstuhl,CollbergBook,desutter2024evaluation}.

Existing work on understanding the relevant aspects of RE has mostly been based on interviews and surveys conducted with reverse engineers~\cite{Ceccato2019,Sutherland2006,Votipka2019}, on knowledge extraction from written reports~\cite{Ceccato2017,Ceccato2019}, and on performance metrics obtained with controlled experiments~\cite{2014afamily,2016comparing,viticchie2016assessment,2014another}. Interviews, surveys, and reports depend on the subjects’ memory accuracy, their answering/reporting style, their willingness to disclose information, and their limited availability for answering/reporting. Furthermore, there is always bias resulting from the choice and formulation of the questions and topics to be addressed.
Such studies hence all have limited precision (in the sense of producing the same results when they are repeated) and completeness, even if they are conducted with established qualitative research methods such as \emph{open coding}~\cite{Flick}.\footnote{This paper is positioned at the intersection of software engineering on the one hand, and qualitative research methodologies on the other. Terms such as `code', `coder', and `coding' have different meanings in those domains. To avoid confusion, we will use the terms `annotation', `annotator', and `annotating' to denote what is know as code, coder, and coding in qualitative research methodologies, reserving those latter terms to denote software, software developer, and software development.}
Finally, the production, collection, and processing of interviews, surveys, and reports is a labor-intensive process, inducing a significant amount of effort on the researchers conducting the studies, as well as reporting overhead on the subjects. As a result, the aforementioned studies are all one-offs. Executing similar studies on a continuous basis is simply not realistic.

Alternatively, automated data collection techniques can be used to gather activity logs during the execution of RE tasks. For example, Mantovani et al.\ acquire activity logs with a web-based interactive disassembler with built-in custom activity logging~\cite{Mantovani2022}, and H\"ansch et al.\ acquire logs of activities in the Eclipse IDE with the Fluorite plugin~\cite{hansch2018programming}. These logs, which fundamentally correspond to a timeline of time-stamped annotations, are then searched for patterns to identify different types of RE activities and strategies. 

\begin{figure}[t]
\centering
\includegraphics[width=1.0\textwidth]{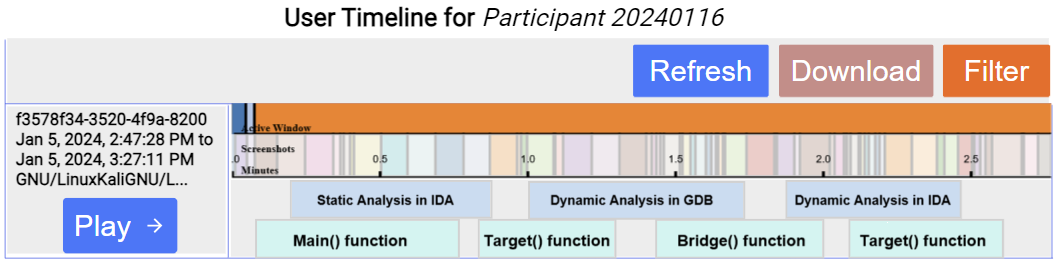}
\caption{\label{fig:timeline} Simulated timeline view, showing different types of annotations, such as those indicating which form of analysis was being deployed, and which functions (main, target, bridge) were being displayed in a tool at which times.}
\end{figure}

Such tool-specific logging approaches to obtain a timeline of annotations do not generalize to (closed-source) third-party RE tools that lack support for activity logging.  Taylor's approach of tool-agnostic activity logging~\cite{Taylor2022,Taylor2019} can complement the tool-specific logging. Her RevEngE framework logs time-stamped screenshots, keystrokes, mouse clicks, active process lists, active window information, etc.\ regardless of the specific RE tools used by the engineers. Part of Taylor's system is a visualization tool in which a timeline of collected data can be viewed by a human annotator. After interpreting the data, the tool allows the analyst to add relevant annotations to the timeline. Figure~\ref{fig:timeline} illustrates this, with annotations expressing which activities were taking place as well as on which artifacts those activities were performed.

Taylor's approach suffers from a lack of scalability, however, because manually identifying the relevant annotations has proven to be too time-consuming to scale to large experiments: In our first experiments with Taylor's tools to annotate collected logs, we found that annotating one minute of RE activity requires more than one minute of a researcher's time. Such annotating needs to be performed by multiple RE researchers for reasons of accuracy and cannot be outsourced to, e.g., a Mechanical Turk because of the complex and domain-specific nature of the data to be annotated. We therefore concluded that in order to scale, Taylor's method needs to be augmented with a semi-automated annotation process.

The main target of the design science research~\cite{hevner2004design} on which we report in this paper is to develop and evaluate such an annotation process and the necessary tool support, with the aim of providing answers to our central research question (RQ):
\begin{quote}
\emph{RQ1: Is it possible to automatically annotate a timeline of data collected on a reverse engineer’s machine with annotations pertinent to how their reverse engineering strategies are executed?}
\end{quote}

This broad RQ encompasses many aspects, so it can be refined into a number of more focused RQs. In the remainder of this paper, we present our evidence and arguments to answer the following RQs:
\begin{itemize}
\setlength{\itemindent}{-.1in}
    \item RQ2: What are the relevant annotations for annotating RE timelines? (Sec.\ 2)
    \item RQ3: What data can we collect on reverse engineers' machines? (Sec.\ 3)
    \item RQ4: What forms of data collection are acceptable for them?  (Sect.\ 4)
    \item RQ5: What is the impact of data collection on their modus operandi? (Sec.\ 5)
    \item RQ6: What annotations can we extract automatically from collected data? (Sec.\ 6)
    \item RQ7: What is the reliability of the extracted annotations? (Sec.\ 7)
\end{itemize}

The evidence and arguments that we will present includes a literature study, informed arguments, usage scenario discussions, a survey we conducted, observations on multiple field studies in the form of RE experiments we conducted with human subjects, and a quality evaluation we performed on various data sets. 

The paper's contributions are:
\begin{itemize}
    \item We present reAnalyst, an analysis framework designed for analysing RE experiments, with a focus on our method for annotating RE activities in tool-agnostic activity logs.
    \item We present adaptations to Taylor's RevEngE framework that were necessary to enable reliable annotation.
    \item We present an evaluation of the reliability of our method to extract useful annotations from tool-agnostic data such as keystrokes, mouse clicks, and screenshots. 
    \item We report on multiple experiments in which we used RevEngE and reAnalyst and a survey we conducted regarding this technology. 
    \item We open source our framework.
\end{itemize}

The remainder of this paper is first structured along the different RQs listed above. After the sections answering those RQs, Sections~\ref{sec:futurework},~\ref{sec:availability}, and ~\ref{sec:conclusions} discuss future work, artifact availability, and conclusions, respectively.

\section{Reverse Engineering Activity Annotations}
RQ2 concerns the types of RE annotations that researchers might be interested in to describe how human experiment subjects execute their RE strategies. In this section, we first study which RE annotations have shown to be relevant in the literature. We then discuss our observations and frame our own work. 

\subsection{Literature Study}
\label{sec:annotations:literature}
Faingnaert et al.~\cite{checkmate2024} discuss that modeling RE activities and the effort required to execute RE strategies requires capturing (i) the \emph{artifacts} that make up the software, such as functions and instructions; (ii) the \emph{relations} between those artifacts, such as control and data dependencies; (iii) the \emph{properties} of artifacts and relations, such as execution frequencies or a basic block's instruction mix; (iv) the \emph{mappings} between concrete artifacts and more abstract ones, such as the fact that a specific sequence of instructions forms an opaque predicate computation; (v) four classes of \emph{activities}, namely data collection, comprehension, localization, and decision-making activities.

\begin{figure}[t]
    \centering
    \includegraphics[width=11cm]{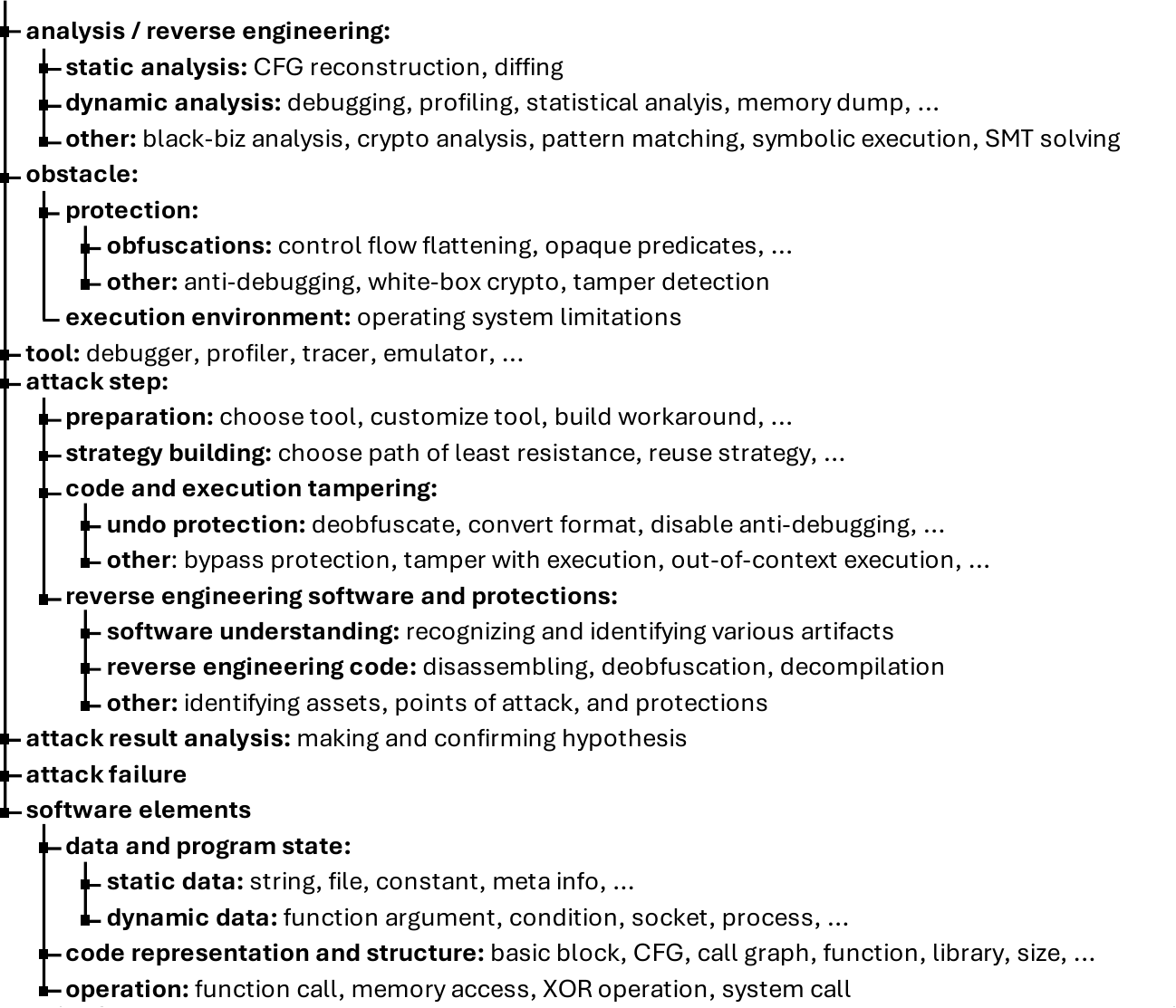}
    \caption{Sample of the taxonomy by Ceccato et al.~\cite{Ceccato2017,Ceccato2019}.}
    \label{fig:ceccato}
\end{figure}

Ceccato et al.~\cite{Ceccato2017,Ceccato2019} construct a taxonomy of annotations in their research on the behavior of software hackers. Their taxonomy is based on reports of professionals and interviews with an amateur regarding their man-at-the-end (MATE) attacks 
on protected software. The annotated reports and interview enabled the construction of  four models of hacker activities relating to (i) obtaining code comprehension, (ii) making and testing hypotheses and building and executing \changed{RE} strategies, (iii) choosing, customising, and creating new tools, and (iv) defeating protections. To illustrate the breadth of their taxonomy, we reproduce a small sample in Figure~\ref{fig:ceccato}. Notice how the concepts in it cover the concepts that Faingnaert et al.\ put forward. Also notice how the artifacts as well as the activities range from concrete to abstract, and from general to domain-specific. 

Numerous human experiments have been presented to evaluate how effectively obfuscations hinder RE tasks such as code localization, code comprehension, and code modification~\cite{exp_eval_obf_against_reveng,
2019impact,
2014afamily,
2008twoardsexperimental,
2009assessment,
2020_creative_manual_code_obfuscation_as_a_countermeasure_against_software_reverse_engineering,
kuang2018enhance,
obf_optvialangmods,
liu2017stochastic,
vmguards,
2020splitting,
viticchie2016assessment,
2019resilient,
2014another}. These papers typically report on total task execution times, as well as task success ratios. This data is typically gathered in post-questionnaires or through direct observation during the controlled experiment. This corresponds to \emph{timestamped annotations ``start'', ``finish'' and ``give up''}. More insightful data could be obtained with more fine-grained annotations. For example, when one obfuscates a program, the obfuscations are typically applied only to the critical code fragments. With only the above three annotations, an analysis cannot discriminate between time spent on obfuscated fragments and time spent on non-obfuscated fragments. In other words, it will not be possible to discriminate between time spent on treated parts of the challenge binaries and time spent on untreated ones. The only measurable effect size of the deployed obfuscation is then that of time spent on treated and on untreated parts combined. In this, the unknown time spent on untreated parts is pure measurement noise. If additional annotations would indicate which code fragments of a program are being studied at which points in time, this would enable measuring the effect size on only the treated parts, thus obtaining a more precise measurement. 

H\"ansch et al.~\cite{hansch2018programming} went further in their analysis of humans \changed{reverse engineering} obfuscated code. They replicated the study by Ceccato et al.~\cite{2014afamily} but extended the measurements by counting the subjects' \emph{uses of Eclipse IDE functionality} to investigate how different obfuscations impact the executed RE strategies. The logged functionality included file open operations, automatic identifier renaming, construction of call graphs and type hierarchies, number of program executions and their total execution time, number of times and total time spent in debugging mode, and amount of time spent reading code. Each entry in the log obviously corresponds to a timestamped annotation. Viticchi\'e et al.~\cite{2020splitting} also went further to investigate the most effective \changed{RE} strategies against different obfuscation configurations: In written post-experiment reports, they annotated in which \emph{functions} the subjects had modified the code in the challenge application. 

Mantovani et al.~\cite{Mantovani2022} also tracked which \emph{functions} and \emph{basic blocks} reverse engineers study, and in what order they visit them as they traverse the challenge binaries.

Some reverse engineers, academics, and practitioners share their knowledge by presenting RE \emph{case studies} in scientific papers and blogs. For example, Guillot and Gazet present a novel RE tool and describe how it improves their productivity on a concrete RE task~\cite{binprot_semiautotamp,guillot2010automatic}, Rolles describes how to customize tools to \emph{reverse engineer} software protected with code virtualization, and details all steps of his \emph{RE strategy}~\cite{deobf_virtualization}, and Biondi and Desclaux describe how they reverse engineered Skype~\cite{desclaux2006silver}. In their descriptions, these reverse engineers refer to \emph{different types of artifacts, relations between them and properties of them}, as well as to the \emph{tool functionality} they used.

To label activities observed in capture-the-flag games (which do not necessarily involve software RE, but can instead involve all kinds of other cybersecurity-related actions) Savin et al.~\cite{Savin2023} used closed annotating based on the 234 attack techniques (and even more sub-techniques) from the \emph{MITRE ATT\&CK Matrix for Enterprise} (\url{https://attack.mitre.org})~\cite{strom2018mitre}. They developed the Pathfinder tool that can assist human annotators by offering them an efficient UI to select entries from the matrix. 

Finally, Wong et al.~\cite{wong2021inside} conducted semi-structured interviews with malware analysts regarding their day-to-day practice. They annotated the responses for, among other, \emph{time spent per malware sample, provenance data of the samples (source, age, family, ...), used sandboxes and their configuration, used software applications and analysis tools, malware signatures and how those were generated}.  

\subsection{Discussion}
The above literature study shows that the answer to RQ2 (``What are the relevant annotations for annotating RE timelines?'') is that there is a wide range of annotations that are relevant to label activities in RE experiments. This includes annotations that refer to concrete artifacts such as specific instructions in a binary, but also annotations that refer to abstract concepts such as ``confirming a hypothesis'' regarding a deployed protection. If the subjects in an experiment have complete freedom-to-operate~\cite{Ceccato2017,Ceccato2019}, open annotating will be necessary in order to obtain the most complete picture. The study also reveals, however, that a small, closed set of annotations, covering a much narrower range of artifacts and/or activities, can already suffice to obtain interesting results from an experiment. 

Which annotations are needed, and whether open or closed annotating should be used, hence depends on the specific focus of the specific RE research being conducted and on the freedom-to-operate of the subjects. In most of the discussed literature, that freedom was severely restricted, because the subjects were students using the tools they had been accustomed to in class, and because the RE tasks were of limited complexity to be doable in a couple of lab hours that fit in a class schedule. Such experiments are definitely the most affordable and \changed{practical to conduct, and hence they are a primary use case for our work.}  

\subsection{Our Work}
With our research, we wish to explore and develop methods to assist researchers for annotating data streams collected during experiments with all of the mentioned types of annotations. We hence do not limit the scope of our work to any particular subset of annotations. We do, however, foresee that this assistance will in many cases not fully automate the annotation process. We hence accept that in some cases, manual post analysis of the collected data will still be needed for annotating the data, as well as for interpreting the annotations. 

Such post analysis can even involve the help of the subjects. For example, we foresee that a researcher may need to conduct post interviews with subjects to obtain additional annotations and to extract knowledge from them, e.g., in the form of models of hacker activities~\cite{Ceccato2017,Ceccato2019}. To prepare pointed questions for such interviews, annotations that have already been attached automatically to the available data can of course be of great help. So even if not all relevant annotations can automatically be determined, we still consider it useful to try to determine as many annotations automatically as possible.

\section{Reverse Engineering Data Collection}

RQ3 concerns the forms of data that can be collected in RE experiments. In this section, we first study which data has been collected according to the literature. We then discuss our possible additional forms, and frame our own contributions (which will be presented and evaluated in detail in later sections) in the existing state-of-the-art.

\subsection{Data Collection In Literature}

In most RE experiments, the researchers collect minimal amounts of data. This includes quantitative data such as (sub)task execution time, (intermediate) result correctness and completeness, and answers to ante and post questionnaires~\cite{exp_eval_obf_against_reveng,
2019impact,
2014afamily,
2008twoardsexperimental,
2009assessment,
kuang2018enhance,
obf_optvialangmods,
liu2017stochastic,
viticchie2016assessment,
2019resilient,
2014another}.

Viticchi\'e et al.~\cite{2020splitting} went a bit further by collecting in which functions the subjects had modified the code of the challenge applications to complete their assignments.  

In some experiments, post-study free-form reports and interview are produced. Sutherland et al.~\cite{Sutherland2006} obtained data from reverse engineers primarily through an exit survey, which collected general feedback about the various stages of the RE experiment. In the experiment by Ceccato et al.~\cite{Ceccato2017,Ceccato2019}, this approach allowed hackers to freely describe their experiences, strategies, and the obstacles they encountered while taking part in the RE experiment. The goal was to capture their unstructured thought processes without restricting them to predetermined questions. However, experiments based on exit surveys and post reports rely heavily on memory accuracy and subjective experiences of subjects, as well as their willingness to disclose complete information, potentially leading to biased, inaccurate, or incomplete data collection~\cite{Nunkoosing2005}. Closed-form surveys also suffer from these issues, and most often will not capture the real-time decision-making processes and detailed strategies of reverse engineers.

Transitioning from this reliance on subject reports, Wong et al.\ and Votipka et al.\ conducted their research using \emph{semi-structured interview} methods. Wong et al.~\cite{wong2021inside} conduct online interviews with professional malware analysts from different companies, aiming to understand their objectives, workflows, and considerations in dynamic analysis system setup. Votipka et al.~\cite{Votipka2019} not only conducted video interviews with subjects to understand their choice of RE techniques but also asked them to show their process while sharing their screens. In other words, this approach  combined interview and observational study methods. While this strategy provides more depth, structured interviews might not adequately capture the nonlinear thought processes inherent to RE.

Others have combined the aforementioned approaches, such as Bryant, who combined a case study with semi-structured interviews and an observational study on how reverse engineers make sense of assembly language representations~\cite{bryant2012understanding}.

H\"ansch et al.~\cite{hansch2018programming} used Eclipse plugins to collect accurate logs of the use of Eclipse functionality during their experiments. Such plugin-based logging, or comparable approaches such as the use of hooks, allows for the most accurate logging of interactions by the subject with the used software. This option is not available for all RE tools, however, and it does require the subjects to install and use researcher-supported versions rather than the versions they might use in their day-to-day working environment. 

Mantovani et al.~\cite{Mantovani2022} developed their own browser-based interactive disassembler that offers only limited functionality, and that produces accurate logs of which \changed{function control flow graphs (CFGs)} and which basic blocks in those CFGs are visited and looked at in which time frames. To obtain accurate logs of which specific basic blocks are being studied, their disassembler at any point in time shows only one basic block in a CFG in full on the display, with the other blocks being blurred. While this allows accurate data collection it can obviously have a large impact on the subject's user experience, and hence puts in question the validity of the study result.

During capture-the-flag games, which do not necessarily involve software RE, but can instead involve all kinds of other cybersecurity-related actions, Savin et al.~\cite{Savin2023} collected keystrokes, time-stamped logs of commands entered in the bash shell, as well as a complete video recording. 

Unlike the works of H\"ansch et al., Mantovani et al., and Savin et al., Taylor's approach of tool-agnostic activity logging~\cite{Taylor2022,Taylor2019} is not limited to specific tools. Her RevEngE client-site software collects screenshots, keystrokes, mouse clicks, active process lists, active window information, etc.\ regardless of the specific RE tools used by the reverse engineers. In addition, the client offers the reverse engineer the option to type in comments during the experiment, with which they can express any thoughts they wish to share with the researchers. 

\subsection{Additional Data to Collect}\label{sec:additional_data}

It is interesting to consider whether the types of RE data data presented in the research literature are exhaustive, or if there are additional types of data that could be collected that would enhance the analysis. For example, given the state-of-the-art in speech-to-text, collecting voice comments instead of the written comments in Taylor's RevEngE would significantly ease the collection of subjects' thoughts. However, this would only be an option in experimental setups where this is acceptable. In a classroom environment, for example, having all participating students speaking out loud would
be disruptive. In a home environment, nearby individuals who are not participating
in the study (and who have not signed any consent agreement approved by an Institutional Review Board (IRB)) could be accidentally recorded, violating their right to privacy. Taken together, voice recording is problematic in any but the most constrained environments.

In addition, when subjects deploy dynamic analyses, it could be interesting to collect all the inputs that the subjects feed to the challenge binaries, including the complete files that configure it or that contain data inputs. Collecting such files will enable the researchers to replay the executed actions.

Similarly, it might be interesting to collect the internal data of tools the reverse engineer deploys on the challenge binaries. For example, interactive disassemblers like IDA, Binary Ninja, and Ghidra store the information they obtain on a binary in a custom database. In essence, their GUIs offer a practical interface to the data in the database, for example to search it and navigate in it. Having accesses to that database during the annotation process can help a researcher interpret the information shown in screenshots.

Finally, it might be interesting to collect eye movement data and electroencephalogram (EEG) brain data (i.e., waves), as has been done in program comprehension research~\cite{ishida2019synchronized}. Eye movement data can be used to determine more accurately which artifacts in a screenshots a subject is studying, and different brain waves reflect different types of mental micro-tasks such as syntax recognition, calculation, memorization, and the result state thereof, such as having achieved a certain level of comprehension. Obviously, this requires specialized equipment.

\subsection{Our Work}
\label{subsec:ourwork_taylor}
Our work builds on the RevEngE data collection framework from Taylor~\cite{Taylor2022,Taylor2019}. 
We present the reAnalyst suite of tools that extract
annotations from the collected raw data and adds these annotations to the data stream.
The automatically generated annotations complement those obtained from subjects'
own comments, plug-ins that log activities directly from RE tools, and manual
annotating. While the latter might produce higher quality data, it can never scale to
long-running experiments with dozens of subjects.

For an extensive discussion of the technical details and inner workings of RevEngE, we refer to the existing literature~\cite{Taylor2022,Taylor2019}; here, we limit discussion to the aspects pertinent to this work's contribution.

Taylor's data collection framework is not screen recording software. Instead, it records time-stamped activity logs comprised of interactions by the reverse engineers with their computer. Screenshots and window data (coordinates, dimensions, and titles) are gathered periodically and upon user input events. Process and thread details, such as running processes' CPU and memory usage, start and end times, and process states, are collected periodically and when screenshots are taken. Mouse inputs record x/y-coordinates upon mouse button presses, and all keystrokes are captured as keyboard inputs. The collected data is uploaded to a server where it is stored in a relational database. 

The data collection framework is designed to operate on most modern operating systems, including MacOS, Windows, and Linux. It can be configured to serve the needs of different kinds of experiments, including preset participant tokens vs.\ sign-up, continuous automatic data upload vs.\ manually triggered uploads, UI vs.\ no UI, etc.

A typical use is that the experiment subjects install and run that framework in a their VM, in which they download the challenge binary and in which they can install and run any tool chain of their choice. The framework's user interface allows reverse engineers to pause data collection whenever they wish, and it provides a text box through which the subjects can express their thought processes and provide comments to aid the later annotating of the collected data.

\begin{figure}[t]
\centering
\includegraphics[width=0.95\textwidth]{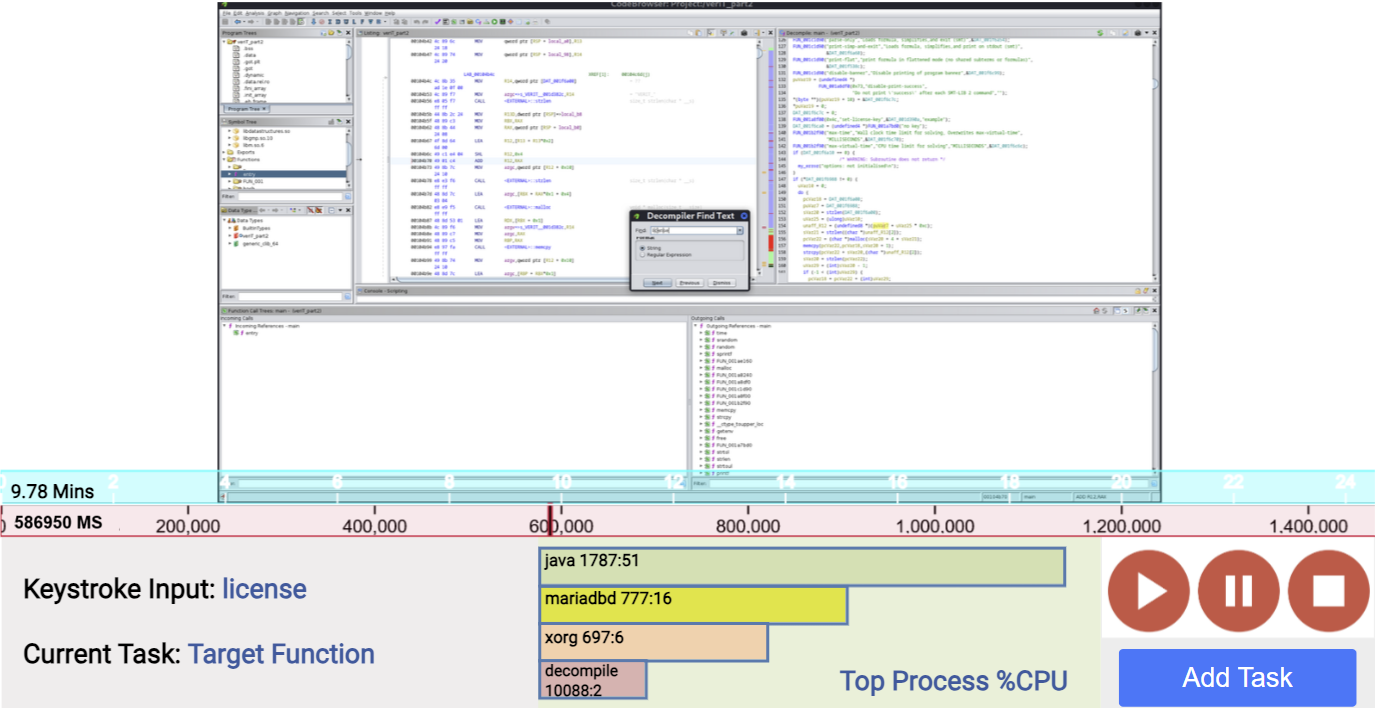}
\caption{\label{fig:timeline-annotations} Simulated animation view, which replays the collected screenshots along with recorded keystroke input (``license''), the current task annotation generated after function annotation (``Target Function''), and the top active processes with their CPU usage percentages. Researchers can use the buttons on the right to play, pause, or manually add task annotations to the timeline.}
\end{figure}

In a web app, researchers can then view or download each session's data via a web visualization or an API, respectively. The visualization displays sessions in a timeline view, complete with annotations as shown in Figure~\ref{fig:timeline}. Additional annotations can be entered in a timeline animation view as shown in Figure~\ref{fig:timeline-annotations}, or headless via the API.

The RevEngE data collection framework as presented in older work~\cite{Taylor2022,Taylor2019} and as used in early experiments did not meet the requirements of the reAnalyst tool that we will present and evaluate in later sections. We hence experimented with alternative design and implementation options. Because high-quality screenshot images are required to achieve good performance in optical character recognition (OCR), we have adapted the screenshot compression scheme to use differential lossless compression~\cite{sayood2017}. We transitioned from the lossy JPG format to the lossless PNG format~\cite{miano1999}, and we capture only the changed regions between consecutive screenshots, which reduces file size and required network bandwidth without compromising image quality. Additionally, we have addressed minor bugs and enhanced the overall user experience.

In addition, we extended the framework to collect tool-specific activity logs from RE tools that allow this. For example, we can now collect Ghidra logs of visited functions. That functionality is orthogonal to capabilities for tool-agnostic logging, which is the core contribution of this paper. We hence only mention this extension for the sake of completeness, but will not further discuss and evaluate it.

\section{Acceptability of Data Collection Methods}
\label{sec:dataacceptance}
RQ3 (``What data can we collect on reverse engineers’ machines?'') relates to the acceptability of data collection methods in the eyes of reverse engineers. In this section, we focus on the automated collection of data during an experiment, with software such as RevEngE, and we present the evidence we gathered regarding its acceptability in the forms of observations on past experiments and a survey we conducted among reverse engineers. This evidence complements the extensive discussion on this topic in Claire Taylor's dissertation~\cite{Taylor2022} and confirms its main message that various types of rewards for participation are the standard way to overcome trust and acceptability issues.

\subsection{Pseudonymous Master Student Experiments}
\label{subsec:master_experiment1}

To test our approach, we conducted experiments over three years with master students enrolled in a Software Hacking and Protection course at Ghent University. To comply with the University's IRB requirements, we ensured that participation was pseudonymous, with unlinkable pseudonyms. 
Prior to installation, students were presented with a detailed consent form. 
Only after agreeing to these terms did the installation proceed. 
Participation was entirely optional. No academic incentives were offered; participants pseudonymously received a small electronic gift card. 

Over the three years, 83 out of 136 students registered to participate, representing an overall participation rate of 61\%. We do not know the reasons for the other 39\% of students choosing not to volunteer. Table~\ref{tab:participation_experiments} presents year-by-year data that shows an increasing participation rate. The data hints at two possible reasons for this increase. First, an increase in gift card value might help explain the 100\% participation in the last year. It might also be, however, that the increase is due to the lecturer (B.\ De Sutter) becoming better at introducing and framing the scientific experiments in the context of the course, leading to a more positive mindset of the students towards those experiments. In the last year, the course became more focused on MATE software protection compared to previous editions, in part by the inclusion of a module that focused specifically on scientific methodology in software protection research.

\begin{table}[t]
    \centering
    \includegraphics[width=0.95\textwidth]{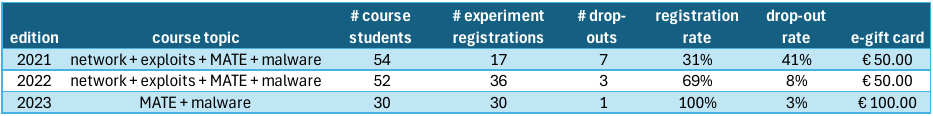}
    \caption{Student participation evolution}
    \label{tab:participation_experiments}
\end{table}

These results suggest that this form of in-class experiment, where participants’ identities are protected and participation is framed and incentivized but optional, is generally acceptable to students studying RE. This approach provides a model for data collection that respects participant privacy while allowing for meaningful engagement.

\subsection{Grand Reverse Engineering Challenge}
\label{subsec:GREC}

In addition to student experiments, we also organized the \emph{Grand Reverse Engineering Challenge} (GREC), a public event open to a broader audience. Unlike the controlled classroom setting, this challenge allowed anyone to participate, with no restrictions on their tools, timing, or approach, provided they enabled our data collection framework. The challenge ran for two months, during which participants could attempt to solve any of the ten RE challenges based on ten unique binaries. Participants who successfully completed challenges were eligible for monetary prizes. More information about this GREC can be found at \url{https://grand-re-challenge.org}~\cite{grand2024challenge}.

To ensure winners received their prizes, participants were required to register with an email address. Each email address can be registered only once. This minimal level of identification was necessary to facilitate prize distribution and manage user accounts. 

In total, 26 user accounts were created for the GREC and provided data from 112 RE sessions. However, it is important to note that this number may include multiple registrations by the same participant (under separate email addresses). This approach differs from the pseudonymous student experiments by adopting a more flexible structure with minimal identification requirements. The challenge demonstrates the feasibility of engaging a broader audience in RE experiments with on-the-fly data collection through this incentive-driven model.

\subsection{Survey for Reverse Engineers}\label{reversesurvey}
To further assess the acceptability of our data collection framework, we conducted an anonymous survey in October--November 2024 targeting professional reverse engineers and others with RE experience. The survey link was shared privately with individuals in our network and distributed at a computer security conference, and we received 23 valid responses. The full list of survey questions is provided in \ref{appendix:survey}.

The survey starts with a brief description of our framework, explaining its functionality and data collection methods. Responses indicated that 22\% of participants are employed by companies that impose restrictions on their involvement in such experiments, meaning they are not allowed or cannot freely participate due to employer policies or contractual obligations. Among the remaining 78\%, who are not restricted by an employer, their responses suggest general acceptability of our framework. 

Among those who are not restricted, when asked about personal policies or ethical considerations that might restrict participation, only 8\% reported significant considerations, while 69\% had no concerns, and 23\% were unsure. Regarding concerns about installing third-party data collection software, 46\% had no concerns, 39\% had minor concerns, and 15\% expressed significant concerns. Additionally, 77\% of respondents indicated that participating in the experiment would not conflict with any contractual obligations, while 23\% were unsure.

The respondents included graduate students (55\%), university faculty (6\%),  company employees/professional hackers (33\%), and hobbyists (6\%). These results suggest that our data collection framework is generally acceptable for most participants, provided there are no employer-imposed restrictions. This reinforces the viability of our framework for broader use in RE research involving professionals, while also highlighting the importance of accommodating potential restrictions from specific employment situations.

\subsection{Ethics and Privacy Discussion}
Data collection frameworks like RevEngE obviously raise ethics and privacy concerns. Its use requires the consent of participants, which in turn requires full transparency about the data collection. The fact that our system is completely open source certainly helps in this regard. The pause button also helps, as it allows participants to interrupt the data collection any time they would want to perform a privacy-sensitive action, such as quickly checking some emails, or entering a password to authenticate on some website. As our framework does not record voice data yet, we also face no risk of accidentally recording data from non-participants without their consent. 

The Ethics Committee of the Faculty of Engineering and Architecture approved the experiments with master students reported in this paper, as well as the implemented procedures as documented at \url{https://github.com/csl-ugent/reAnalyst}. Full anonymity (i.e., pseudonymity with unlinkable pseudonyms) needs to be guaranteed in such experiments when the researcher leading the experiment is in a position of power over the participants. This is the case, for example, if the leading researcher is the professor responsible for teaching and grading the students in software protection or RE courses. Without full anonymity, it is impossible for the students to give their free consent. 

The Institutional Review Board (IRB) of the University of Arizona approved the GREC, ensuring compliance with ethical standards for public data collection. The fully anonymous survey reported in Section~\ref{subsec:survey} aligns with the ethical stipulations of our universities and, as such, received a waiver for conducting it.

\section{Impact of Automated Data Collection On Reverse Engineering Behavior}\label{sec:dataimpact}

RQ4 (``What forms of data collection are acceptable for [reverse engineers]?") concerns the impact that running data collection software might have on the behavior of the reverse engineers being monitored with it and on their analysis tools.

First, however, we want to highlight that in almost all of the RE experiments with human subjects discussed in the literature, the researchers designing the experiments also designed the challenge binaries. The researchers conducting an experiment hence almost always  control which analysis evasion techniques are implemented in the challenge binaries, including whether that software has dormant functionality that will not be executed if it detects being executed in a VM, in some other sandbox or emulator, or on a system running our data collection software.

The only major exception are experiments in which malware analysts study real malware samples. Malware is known to use dynamic analysis evasion techniques~\cite{malware_evasion}, and malware analysts adapt their practice accordingly~\cite{wong2021inside}. In case malware analysis experiments are conducted using our software additional countermeasures might be necessary. These are out of scope of our work, however, as we focus on experiments in which the researchers create the challenge binaries. 

\subsection{Pseudonymous Master Student Experiments}
\label{subsec:master_experiment2}
Our current data collection framework is designed to have minimal or no impact on the performance of VMs and the tools running in them, assuming participants follow the recommended VM settings we provide. Under our Bring Your Own Device (BYOD) policy, participants use their own machines for experiments. We recommend a VM configuration with 8GB of memory and 3 CPU cores. 

In our 2021 experiment, which served as a trial, several participants encountered performance issues, leading some to discontinue their participation. During this experiment, a significant number of students reported that the data collection framework noticeably slowed down their VMs. Consequently, some students filled out a stop participation form requesting data deletion due to technical difficulties. In total, 7 out of 17 students (41.2\%) discontinued in 2021, 3 out of 36 (8.3\%) in 2022, and only 1 out of 30 (3.3\%) in 2023, due to technical or other personal reasons.

In the 2022 and 2023 editions, we conducted a survey asking students if they experienced any issues with the data collection framework. In 2022, none of the students reported VM slowdowns, and in 2023, only one student (3.3\%) reported a performance issue. However, the specific cause of this reported issue is unclear, as the student did not provide additional details. These results suggest that, with the recommended settings, the framework’s performance impact has been consistently low.

\subsection{Survey for Reverse Engineers}
\label{subsec:survey}
In our survey introduced in Section \ref{reversesurvey}, we also asked participants how the presence of an automated data collection framework might impact their RE strategies or choice of tools. This question aimed to understand whether the framework’s presence would influence their typical RE behavior and potentially introduce bias.

When asked,\textit{``Would the presence of the data collection framework cause you to alter your usual RE strategies or tools?''}, 38\% indicated that simply knowing they were being observed might unintentionally affect their RE approach. Another 15\% cited other reasons for potentially changing their behavior. In contrast, 47\% stated that the framework’s presence would not impact their usual RE methods.

These responses highlight a potential source of bias, known as the ``observer effect,'' where subjects alter their natural behavior due to awareness of being monitored. This effect, documented across various domains, can influence how subjects approach their tasks~\cite{kalvemark2022observer, oswald2016hawthorne}. Such biases are not unique to our method; similar potential biases can arise in RE studies using semi-structured observational techniques~\cite{wong2021inside, Votipka2019}, for example. This effect is important to consider when interpreting data collected under observation, as it may limit the generalizability of the results.

\section{Automated Annotation Extraction}\label{sec:knowledge}
To provide a first answer to RQ6 (``What annotations can we extract automatically from collected data?''), we designed and implemented an initial set of tools to extract time-stamped annotations from tool-agnostic data, i.e., to complement tool-specific logs. 

The raw data collected by RevEngE contains a wealth of information
about the subject's RE activities, and our current system only analyses
a fraction of what might be possible. The comprehensive and open-ended nature of the data
collection ensures that --should future research questions warrant it--- the information
necessary will be available.
In any case, we foresee that our tools can assist researchers in several different ways. For example, the automatically extracted annotations can serve as:
\begin{itemize}
    \item suggestions for annotations from which the researcher can make manual selections, similar to the Pathfinder tool by Savin et al.~\cite{Savin2023} (see Section~\ref{sec:annotations:literature});
    \item the final result of the annotation process, with or without manual post-processing of the generated annotations;
    \item a stepping stone for further manual or automated processes to generate higher-level representations of RE activities. 
\end{itemize}

With our initial analysis tool suite, we focus on automatically deriving the following types of annotations:
\begin{itemize}
\item From time $T_1$ to $T_2$, the subject employed Ghidra/IDA\footnote{We consider multiple disassemblers in our research in order to demonstrate general applicability of our methods to extract information from screenshots, independently of the tools' additional logging capabililities.} whose user GUI showed the control
   flow graph (CFG) for function $F$.
\item From time $T_1$ to $T_2$, the subject employed GDB on binary $B$ to perform a dynamic analysis.
\item At time $T$, the subject employed IDA to rename symbol\footnote{In binary executable terminology, symbols are symbolic identifiers of artifacts in binary executable files~\cite{Levine}. Examples are section names, function names, global variable names, etc. We use the term more broadly here, as it can also denote any other construct that is given a name by the reverse engineers or their tools during the RE. A good example is a globally allocated string in a binary. Disassemblers typically replace string addresses with string identifiers because that eases human code readability. Another example of what we consider symbols are mnemonics of disassembled instructions, and register names occurring in the assembly code obtained by disassemblers.} $Y$ to $Z$. 
\item At time $T$, the subject employed Ghidra to navigated from global variable $V$ to function $F$ via a cross-reference. 
\end{itemize}

This section presents our tool design and, where relevant, implementation aspects, as well as two demonstrations. Section~\ref{sec:evaluation} will evaluate the quality of the extracted annotations.

\subsection{Data Pre-processing}\label{subsec:datapreprocessing}
Before annotating the collected data streams our tools pre-process the data to make it more suitable for the actual annotation tasks.

From the screenshots, we extract the displayed text with the open-source OCR tool Tesseract~\cite{smith2007overview} and with the commercial Abbyy FineReader.\footnote{
We found that Tesseract works as well on high fidelity screenshots, but that ABBYY performs
better on screenshots with lossy compression. While our current system generates PNG screenshots,
some of our early experiments used lower quality JPG images. Abbyy is not free, not open source, does not offer a headless operation mode, and is much slower than Tesseract. We therefore restrict our use of Abbyy to lower quality JPG images.}

To improve the quality of the text recognition and to minimize accuracy issues identified in literature for OCR of screenshots~\cite{Malkadi2020}, we use page segmentation, character whitelisting, thresholding methods, and we upscale the images with a factor 2 before OCR.

Furthermore, our tools use OpenCV~\cite{Bradski2000} to identify contours and bounding rectangles in the screenshots. The relevant rectangles (such as the ones in which basic blocks in CFGs are displayed) are identified based on predefined size and color filters.  Our tools are configurable, should the subject use a custom color scheme.

To identify on which text on screen the reverse engineer clicks, we rely on text bounding boxes obtained with the OCR.

\begin{figure}[t]
\centering
\includegraphics[width=0.4\textwidth]{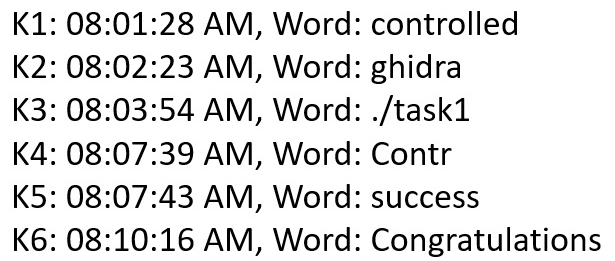}
\caption{\label{fig:keystrokes} Sample keystroke data snippet showing combined individual keystrokes into keyboard inputs.}
\end{figure}

RevEngE collects raw sequences of keystrokes. To be useful, these need to be combined
into meaningful words. reAnalyst combines consecutive keystrokes that are likely part of the same input and filters out redundant or non-meaningful entries, such as pressing the Windows key or deleted characters. The result is a list of time-stamped inputs, as illustrated in Figure~\ref{fig:keystrokes}. 

When a researcher notes that a subject in an experiment uses custom aliases
for shell commands or has renamed some binaries, the researcher can reconfigure
tools that extract data from the keystroke logs and from the process data.

Of this data pre-processing, OCR by far requires the most computing resources. With Tesseract, the average processing time for OCR is 4.8s per screenshot in our experimental environment (on which Section~\ref{setup} will provide details). This includes the time it takes for upscaling images. OpenCV requires much less time, at 0.05s per screenshot. The pre-processing of keystrokes requires negligible resources, at less than a second per whole RE session.

\subsection{Challenge Binary Artifacts}\label{subsub:metadata}
In order to identify a challenge binary's artifacts (such as basic blocks and instructions) that are shown in screenshots and the activities that the subjects perform on them (such as clicking on a cross-reference to navigate from a code fragment to a referenced global variable), our annotation tools need to have a priori knowledge of the artifacts that actually make up the binary executable. In line with the discussion in Section~\ref{sec:additional_data}, our tools collect that information by importing it from the custom databases that tools such as IDA Pro, Binary Ninja, and Ghidra build. This includes the binary's functions, each function's basic blocks, global variables and strings, cross-references (such as which basic blocks contain references to strings or other global variables), and the symbolic names that the tools assign to those artifacts (such as assigning the name FUN\_0x1234a0 to the function at offset 0x1234a0). Commonly occurring, non-discriminative symbols, such as assembly instructions, registers, and other predefined elements, are filtered out. The remaining symbols are further sanitized by removing patterns such as tool-specific prefixes. 
The output of this process are maps from symbol to function and from symbol to basic block.

When unstripped versions of the binaries are available, our tools can extract the original
symbol names. This allows more researcher-friendly annotations such as ``navigates to function
QuickSort'' rather than ``navigates to function FUN\_0x1234a0''.

\subsection{Generating Annotations}\label{subsec:gencode}

From the pre-processed data and the data describing the challenge binary artifacts, reAnalyst can generate many different types of annotations. We describe these next.

\subsubsection{Displayed Function Annotations}\label{subsub:function_matching}
Annotations that show which function is being displayed in a tool can be used to evaluate the effectiveness and efficiency of different RE strategies, as done by, e.g., Mantovani et al.~\cite{Mantovani2022}. 
To generate such annotations from the screenshots, reAnalyst first maps each screenshot to the function(s) shown in it, using information from the
symbol-to-function map and OCR outputs. The process assigns each symbol in the screenshot to a specific function using a combination of regular expression and fuzzy matching techniques, for which we use the FuzzyWuzzy library~\cite{Hall1980ApproximateSM}. Fuzzy matching calculates a similarity score between extracted symbols and the entries in the symbol-function map. If a match with a perfect score of 100 is found, the tool immediately associates the screenshot with that function. For less certain matches, the highest-scoring symbol above a user-adjustable threshold determines the function. The time to process this mapping depends on the complexity of the screenshots: in our environment (see Section~\ref{setup}) it takes 1.7s per screenshot on average.

From sequences of screenshots and their associated functions, different annotations can then be generated according to the preferences of the researcher. For example, for a sequence of screenshots associated to the same function, annotations such as ``main() function'' with a start-end interval can be generated, or annotations such as ``entered main() function'' that only have one time-stamp. reAnalyst can be configured to exclude short intervals, which occur when a function is visible for only a few seconds, such as when the subject is scrolling around. \changed{The minimum threshold is user-adjustable. However, researchers may also opt to keep these intervals to track potential scrolling or quick navigation jumps, which can provide insights into how a reverse engineer scans and revisits parts of a binary.} Minor gaps in screenshot sequences can also be ignored, as these may result from OCR inaccuracies,
function identification errors, or when the subject briefly switches to another window. 

\begin{figure}[t]
\centering
\includegraphics[width=1\textwidth]{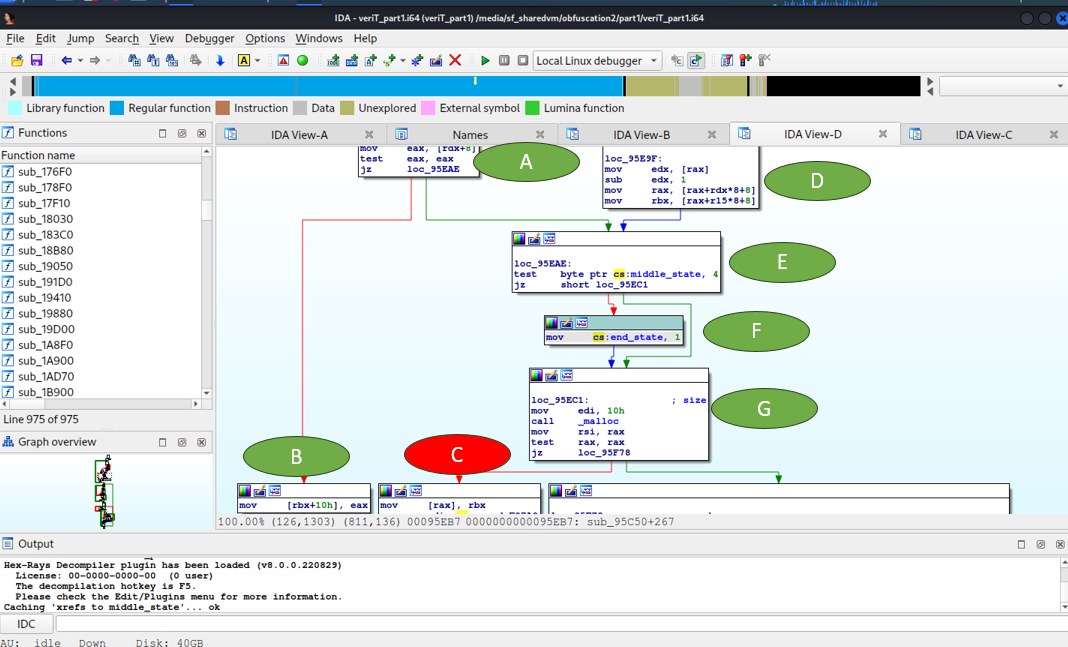}
\caption{\label{fig:cropped_basic_blocks}  \changed{An example screenshot of IDA Pro showing seven basic blocks marked A-G of a function’s CFG. Some of the blocks are only shown partially. Still, the framework identified all but basic block C correctly, which has only a minimal portion visible.}}
\end{figure}

\subsubsection{Displayed Basic Block Annotations}\label{subsub:basic_block_matching}
Just as for functions, reAnalyst generates annotations for basic blocks shown on screen by disassemblers. \changed{Figure~\ref{fig:cropped_basic_blocks} shows an example screenshot that contains a partial CFG, with only a few blocks shown. Annotating which basic blocks are being displayed allows precise tracking of user interactions with the CFG.} 

To match the basic blocks shown on screen to the corresponding basic blocks in the binary, reAnalyst
compares the text extracted from a basic block contour to the candidate basic blocks within
the corresponding function. We use the Levenshtein distance algorithm~\cite{levenshtein1966binary} to find the best match. The average processing time in our environment is 0.03 seconds per basic block.

\subsubsection{Navigation Annotations}
Navigation through the challenge binary and its representation in a disassembler is an important RE activity. 
To generate navigation annotations, reAnalyst captures interactions with displayed artifacts from the binaries through mouse clicks as well as through various search and navigation windows. Clicking-based navigation involves single or double clicks on displayed symbols and displayed symbolic cross-references to navigate between artifacts. For example, next to global variable locations, Ghidra shows a list of all code locations that reference the global variable. By clicking on such a cross-reference, the user can navigate from the variable to that location. Navigation features include ``Find String'' or ``Search for Functions'' that can be accessed through drop-down or pop-up menus, as well as through shortcut key-bindings. 

For click-based navigation, reAnalyst identifies the specific text tokens being clicked as described in Section~\ref{subsec:datapreprocessing}. For the window-based search functionality, reAnalyst correlates screenshots with preprocessed keystroke data. Screenshots provide the visual context of the features so that reAnalyst identifies which feature is used, while preprocessed keystroke data reveals the associated text input. These steps require negligible resources at less than a second for a 10-minute RE session.

\subsubsection{Artifact Renaming and Retyping Annotations}
When reverse engineers add information to the disassembler's representation of the challenge binary or overwrite information in it, they typically do so to encode knowledge they already obtained about the binary, i.e., to label artifacts in the binary with that knowledge. Common actions include renaming a meaningless
symbol FUN\_0x1234 to ``hashtableInsert'' or changing the type of a global variable from UNDEFINED to INT64. Such actions often follow successful comprehension efforts: the reverse engineer almost literally makes note of their newly acquired knowledge. Annotating such actions is therefore important.

After subjects rename a particular artifact (i.e., its symbol), its new name will be displayed instead of its old one. To correctly interpret the symbol occurring in later screenshots, reAnalyst needs to be aware of such renaming actions. This is another reason for annotating such actions. 

reAnalyst captures such actions by correlating screenshots with keystroke data. For instance, if a tool's Rename Window is visible in the screenshots and keystroke data indicates that ``targe'' was typed during the same period, reAnalyst generates an annotation for the renaming of the function to ``target.''

\subsection{Demonstration}

This section demonstrates how annotations generated with the presented tools can reveal interesting aspects of deployed RE strategies.

\begin{figure}[t]
  \centering
  \small
    \begin{verbatim}
        14:37:48, Symbol: FUN_0010ed40, double click
        14:37:48, Entered Function: FUN_0010ed40
        14:38:09 - 14:37:57: Feature: Rename Function, Word: main
        14:39:52, Symbol: DAT_00288bb, single click
        14:39:54 - 14:40:02: Feature: Edit Label, Word: keyplus0x1000
        14:40:10, Symbol: keyplus0x1000, single click
        14:40:13, Symbol: Find References to keyplusOxl000, single click
        14:40:15 - 14:40:18: Feature: References to
        14:40:26 Entered Function: FUN_001A3A20
        14:40:26, Symbol: bVar8, single click
        14:40:28 - 14:40:29: Rename local variable, Word: license key
    \end{verbatim}
    \caption{\label{fig:sample_output} Annotations extracted from a Ghidra user solving a simple license check challenge.}
\end{figure}

First, Figure~\ref{fig:sample_output} shows the annotations generated by reAnalyst from a ``locate the license key'' challenge~\cite{checkmate2024}. Each entry corresponds to a timestamped activity. For example, from the data in the figure we can
deduce that at 14:37:48 the subject first double-clicked on FUN\_0010ed40 to navigate to it, and then renamed it to ``main.''
At 14:39:52 the subject renames the symbol DAT\_00288bb to ``key\_plus0x1000'', indicating that they identified a critical artifact for the assignment. At \texttt{14:40:13}, the subject uses a "Find References" feature to locate occurrences of \texttt{keyplus\_0x1000} in the challenge binary. This step is crucial for understanding how and where this data is used within the code. Finally, at \texttt{14:40:26} the subject navigates to \texttt{FUN\_001A3A20} and renames the local variable \texttt{bVar8} to \texttt{license key}, at which point the license key was finally found.

\begin{figure}[t]
\centering
\includegraphics[width=0.7\textwidth]{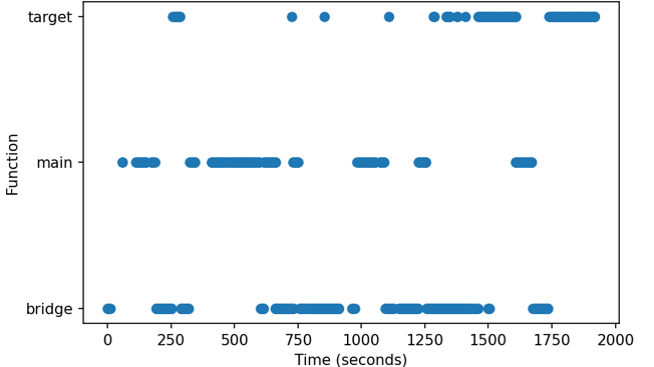}
\caption{\label{fig:scatter} Scatter plot example from function matching data. }
\end{figure}

In a second reAnalyst experiment, we replicated an experiment by Mantovani et al.~\cite{Mantovani2022} that studies reverse engineers' call graph traversal strategies. 
Whereas they had to used a custom disassembler with tool-specific logging developed specifically for their experiment, we rely on annotation generation from tool-agnostic data and our subjects used a third-party disassembler. 
Just like Mantovani et al.\ did from their tool's logs, we generated the chart shown in Figure~\ref{fig:scatter} from the function annotations. 
The chart shows which function in the challenge binary (reused from Mantovani et al.) is being visited at each point in time. 
These types of visualizations can reveal interesting patterns of RE behavior. 
But our data can also be analysed with simple scripts, e.g., to determine whether subjects employed a breadth-first or depth-first traversal strategy over the binary's call graph, or to find correlations between traversal algorithms and RE efficiency or effectiveness. 
This replication experiment confirmed the result by Mantovani et al., namely that no strategy outperforms the other on average. 
Our replication has a lower threat to external validity, however, because our subjects used real-world RE tools rather than Mantovani et al.'s custom lab tool that lacked key navigation features such as string search or navigation by means of cross-references.

These use cases demonstrate how the extracted annotations allow researchers to get a clear view of a subject's navigation through the artifacts of a challenge binary while trying to localize specific artifacts.

\section{Accuracy and Evaluation}\label{sec:evaluation}
To provide an answer to RQ7 (``What is the reliability of the extracted annotations?''), we focus on the most error-prone part of reAnalyst: how to identify displayed functions and basic blocks based on OCR'ed text extracted from screenshots. Such text extraction is prone to errors, which has been observed in previous work
extracting source code from screenshots~\cite{Malkadi2020}.

To evaluate the reliability of reAnalysts' function and basic block identification algorithms, we randomly selected screenshots from a number of data sets collected in experiments with human subjects. These experiments involved a total of 64 subjects with varying levels of RE experience, working on 10 different binaries using RE tools that include IDA, Binary Ninja, and Ghidra. 

Even though a tool such as Ghidra also can support tool-specific activity logging, we include screenshot-based annotation extraction for Ghidra to ensure that our evaluation covers multiple tools with different visualization styles and color schemes.

\subsection{Evaluation Setup}\label{setup}
Our pipeline uses Python 3.10.12 and Java OpenJDK 11, along with OCR tools Tesseract 5.3.4 and Abbyy 16.0. The processing tools operate on an Ubuntu 22.04.4 LTS desktop featuring an Intel Core i7-3770 CPU at 3.40GHz with 8 cores, 16 GB of RAM, and a 480 GB SSD with 47 GB available. Due to compatibility requirements, Abbyy is deployed within a VirtualBox Windows 10 VM on our desktop.

\subsection{Datasets}
\label{sec:datasets}
Our evaluation uses three datasets: Datasets A and B are from the student experiments (see  Sections~\ref{subsec:master_experiment1} and~\ref{subsec:master_experiment2}) and Dataset C is from the GREC (Section~\ref{subsec:GREC}). \changed{We describe details of these datasets in \ref{appendix:datasets}. Table~\ref{table:binaries} lists some features of the challenges in the datasets that are relevant to the remainder of the paper. }

\begin{table}[t]
\centering
\small
\begin{tabular}{ccc}
Dataset               & Binaries & Functions Per Binary\\
  \hline

A & 4 & 961/970/971/967 \\
B & 4 & 977/981/985/981 \\
C  & 2 & 23/112   \\
\end{tabular}
\caption{\label{table:binaries}Overview of challenge binaries in the datasets.}
\end{table}

\subsection{Data Sampling}
To evaluate the accuracy of our methods among diverse users and binaries, we used a stratified random sampling approach~\cite{Parsons2017}. 
From both student experiments (Datasets A and B), from pools of 33 and 29 subjects, 2 individuals were randomly selected. From the remaining subjects, we randomly select a Part 2 session of 1 individual, resulting in a total of 5 subjects and 4 binaries for each dataset. For every selected subject, we randomly selected 100 screenshots, meaning 500 screenshots per dataset. For the GREC (Dataset C), we relied on the data from the sole participant who effectively used our data collection framework across 2 solved challenge binaries. From this subject's sessions, we randomly selected 500 screenshots, for a total of  1,500 screenshots. Table~\ref{table:overview_sampling} summarize the data sampling methodology. While the majority of these screenshots contain functions, some do not, which assesses our framework’s ability to discern screenshots with functions from ones without functions. The datasets used for this evaluation were selected only after the tuning and design of the framework were completed. A set of separate, non-overlapping samples were used during the tuning and design phases to ensure the independence and integrity of our evaluation.

\begin{table}[t]
\centering\small
\begin{tabular}{cccccc}
Dataset & Used RE Tool & \multicolumn{2}{c}{Participants} &\multicolumn{2}{c}{Screenshots}\\    
& & Total & Selected & Total & Selected\\
  \hline
A & IDA & 33 & 5 & 406682  & 500 \\
B & Ghidra & 29 & 5 & 321328 & 500 \\
C  & Binary Ninja & 2  & 1 & 45460 & 500
\end{tabular}%
\caption{\label{table:overview_sampling}Overview of data sampling for each dataset.}
\end{table}

To evaluate the accuracy of the basic block annotation generation, we needed to obtain a dataset of screenshots with a CFG view. All subjects of Dataset A spent a large fraction of their time in graph view browsing in IDA, but the subjects of Datasets B and C spent most of their time on decompiler views. We assume this is mainly because Binary Ninja and Ghidra both offer a decompiled code view, while the free version of IDA that was used for Dataset A does not. We hence only use Dataset A for this evaluation. 
 
Because the accuracy of the basic block annotation depends on the success of the preceding function identification, we used the same screenshots initially selected for function annotations. Due to the tedious process of establishing ground truth for basic block annotations, and because most screenshots contain multiple basic blocks, we chose to work with a subset of those screenshots. From the 100 screenshots already selected from each of the five subjects in Dataset A, we randomly selected 20 screenshots per subject, giving a total of 100 screenshots for this experiment. Screenshots with incorrect function annotation or those not displayed in IDA’s graphical view mode were excluded and replaced with randomly selected alternatives. On average, each of the 100 selected screenshots contains 5.19 basic blocks, leading to a total of 519 basic blocks for this evaluation.

\subsection{Methodology}\label{subsec:methodology}
For all selected screenshots, we compare the generated function and basic block annotations to the ground truth results. The latter were established through manual inspection of each screenshot. Any functions or basic blocks that are identifiable in a screenshot are included in our ground truth results. 

For generating the OCR outputs, we use Tesseract for datasets A and B and for additional OCR processing required for basic block annotations. We use Abbyy for dataset C (which was collected with lossy image compression). Unlike datasets A and B, dataset C was collected using lossy image compression, and our evaluation shows that the image quality was too low for Tesseract to produce acceptable results. We changed from lossy to lossless compression only after dataset C was collected and we realized the issues with it (see Section~\ref{subsec:ourwork_taylor}). 

For the function annotation generation evaluation, we record the result of each annotation to ground truth comparison as {\it Correct Label}, {\it Wrong Function} (when an incorrect function is identified), or {\it No Function}. A ``no function'' is recorded when any part of a function's code is present in a screenshot, but the framework identifies it as a non-function screenshot.  This determination is made solely on the presence of function-related code, regardless of how much of that code is displayed or hidden. This objective approach ensures consistent and unbiased analysis. Additionally, if the screenshot does not contain a function, the outcomes are recorded as either {\it Correct Label} (when the framework correctly identifies it as a non-function screenshot) or {\it Detected Function} (when there is no function in the screenshot but the framework detects one). 

To evaluate the basic block annotation generation, we compare a list of expected basic blocks against detected ones for each selected screenshot. We calculate the number of correctly and incorrectly detected blocks, as well as the number of undetected ones.

\begin{table}[t]
\resizebox{\columnwidth}{!}{ 
\begin{tabular}{c|cccc|ccc|c}
\textbf{Dataset} & \multicolumn{4}{c|}{\textbf{Screenshots have functions}} & \multicolumn{3}{c|}{\textbf{Screenshots have no function}} & \textbf{Overall} \\ 
                 &       & \textbf{Correct}  & \textbf{Wrong}  & \textbf{No}  &  & \textbf{Correct}  & \textbf{Detected}  & \textbf{Accuracy} \\
                 & \textbf{Total} & \textbf{Label} & \textbf{Function}  & \textbf{Function}  & \textbf{Total} & \textbf{Label} & \textbf{Function} & \\
\hline
A     & 362  & 360  & 1  & 1  & 138 & 135 & 3 & 99.0\% \\
B     & 416  & 405  & 7  & 4  & 84  & 82  & 2 & 97.4\% \\
C     & 343  & 328  & 6  & 9  & 157 & 156 & 1 & 96.8\% \\
\hline
Total & 1121 & 1093 & 14 & 14 & 379 & 373 & 6 & 97.7\% \\
\end{tabular}%
}
\caption{\label{table:function_matching_results}Results of function annotations across three datasets.}
\end{table}

\subsection{Results for Function Annotation Generation}
\label{subsec:function_results}

The results in Table~\ref{table:function_matching_results} 
show that our function annotation generator achieves a high accuracy, at 97.7\% overall. Table~\ref{table:function_matching_results} details the function annotation results across the three  datasets. 
The overall accuracy rate is calculated as the ratio of the number of correct labels to the total number of screenshots. 

Further analysis revealed that the performance of our framework is fundamentally dependent on two key factors: the quantity of information within the screenshots, and the accuracy of the OCR output. When these two aspects vary, which they sometimes do given the diverse nature of screenshots from different RE tools, the effectiveness of our framework is put to the test. 

\begin{figure}[t]
\centering
\includegraphics[width=1\textwidth]{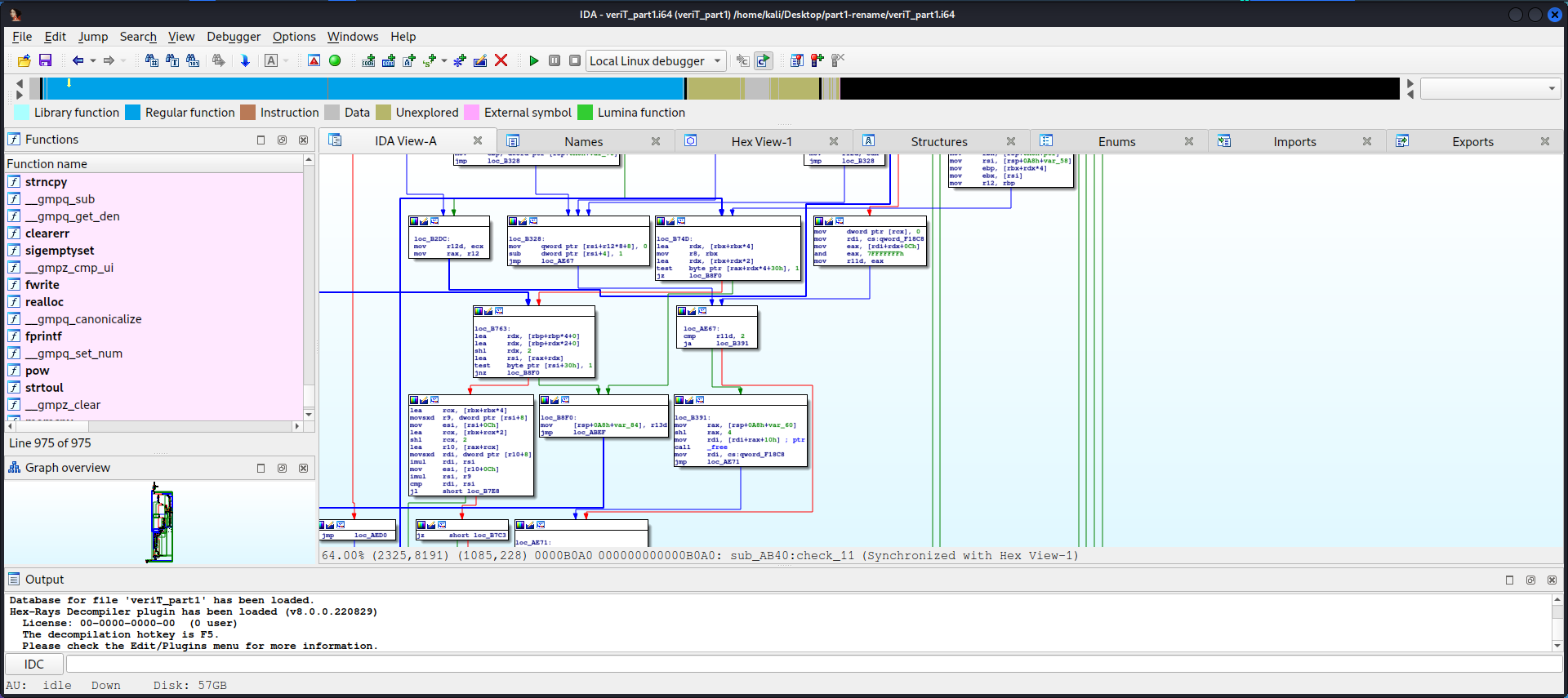}
\caption{\label{fig:zoomed_out_screenshot} The screenshot is zoomed out with a complex graph layout. \changed{Despite the limited character clarity of screenshots like this one, our framework can often extract enough symbols to identify the displayed function.}}
\end{figure}

In general, our framework demonstrated a high degree of resilience in handling a diverse range of screenshots of varying complexity. Generally, it successfully detected and matched functions, even when they were only partially visible or presented in complex layouts. For instance, Figure \ref{fig:zoomed_out_screenshot} displays a zoomed-out CFG, in which the text appears much smaller, which reduces clarity and leads to fewer symbols being recognized accurately by OCR.\footnote{It is worth noting that a tool-specific feature, IDA's display of the current function name below the CFG, is not considered for function annotations in our approach, as it is not unique to a function. Indeed, function names can be shown on screen when the user is watching different functions. For example, a function name can occur in the function list on the left together with other function names, and it can occur in other function's displayed CFGs, namely at its callsites. This current function name shown below the CFG hence does not compensate for the graph being zoomed out or partially obscured.} 
However, it also presents a larger overview of the code, providing more potential symbols to work with.

\begin{figure}[h]
\centering
\includegraphics[width=1\textwidth]{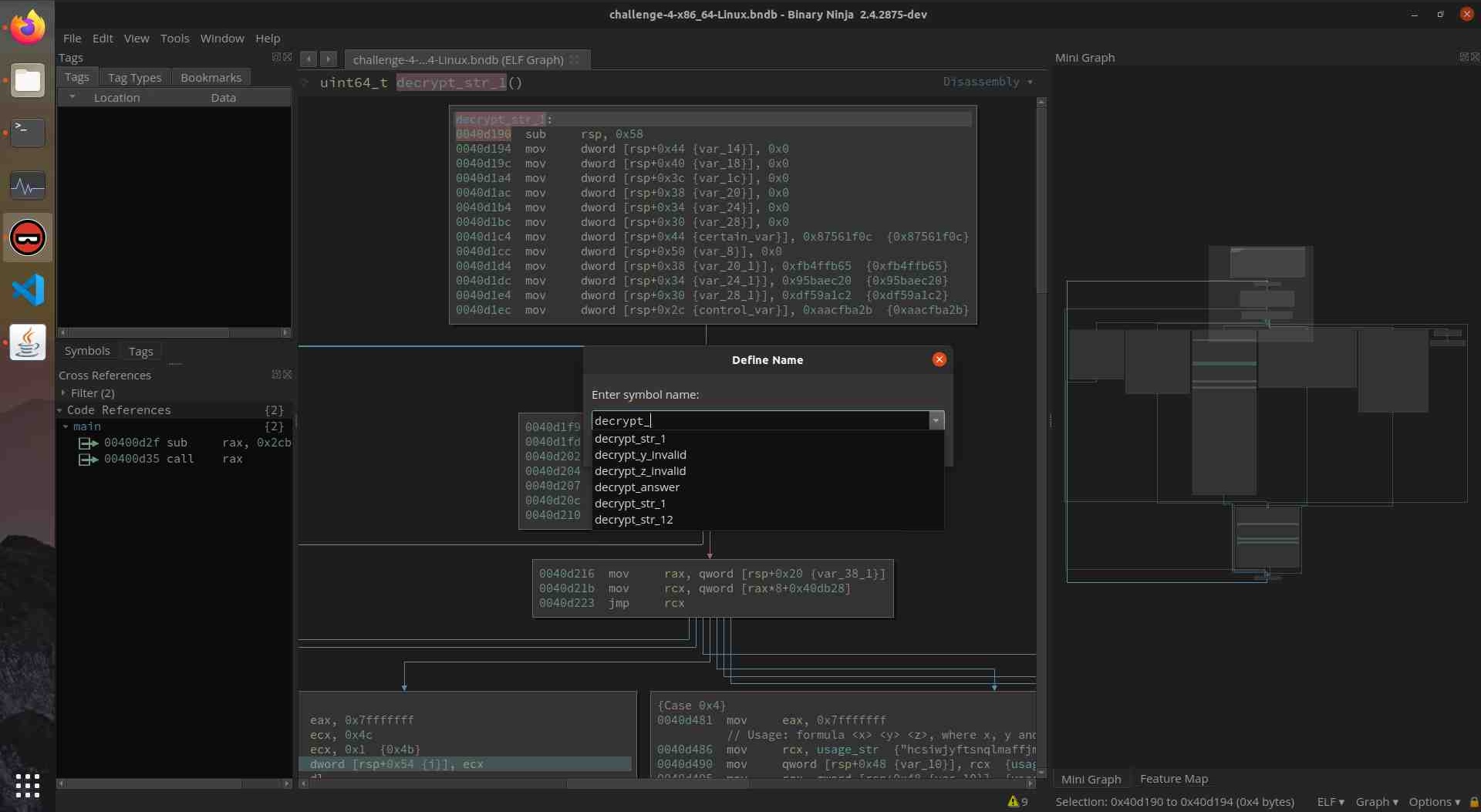}
\caption{\label{fig:obscured_screenshot} The window obscures a portion of the screenshot and makes the interface darker\changed{, but despite these challenges our framework still correctly identified the shown function.}}
\end{figure}

Figure~\ref{fig:obscured_screenshot}, taken from Binary Ninja, features a dark interface and an overlaying ``Define Name'' pop-up window. OCR algorithms are optimized for high contrast between text and background~\cite{Sporici2020}, so a dark interface may affect OCR effectiveness. The pop-up window obscures part of the CFG, which reduces the amount of available information available for function annotations. Despite these issues, our framework still correctly identified the functions using the limited symbols it could extract from images like these.

\begin{figure}[h]
\centering
\includegraphics[width=1\textwidth]{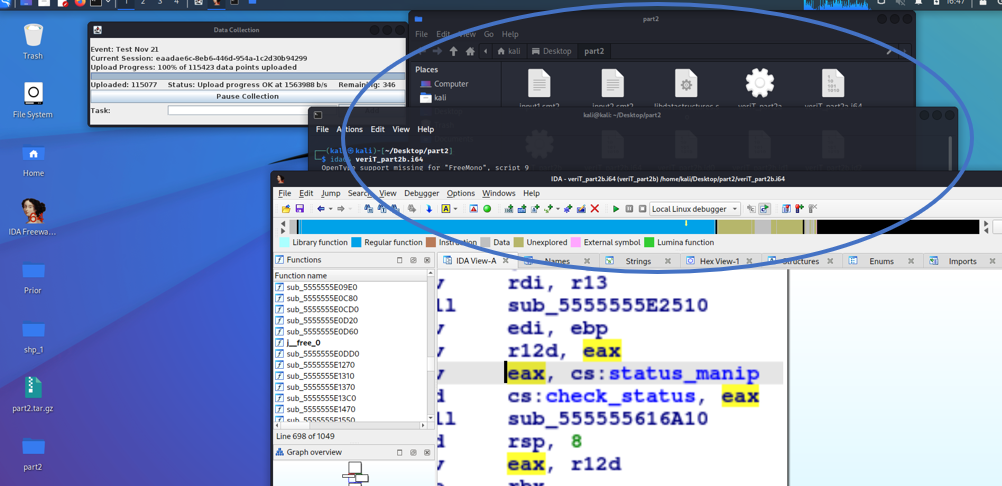}
\caption{\label{fig:false_negative_IDA} The user opened multiple window and only a small portion of the code is visible. \changed{Screenshots like this do not contain sufficient relevant symbols information for a correct function match.}}
\end{figure}

\begin{figure}[t]
\centering
\includegraphics[width=1\textwidth]{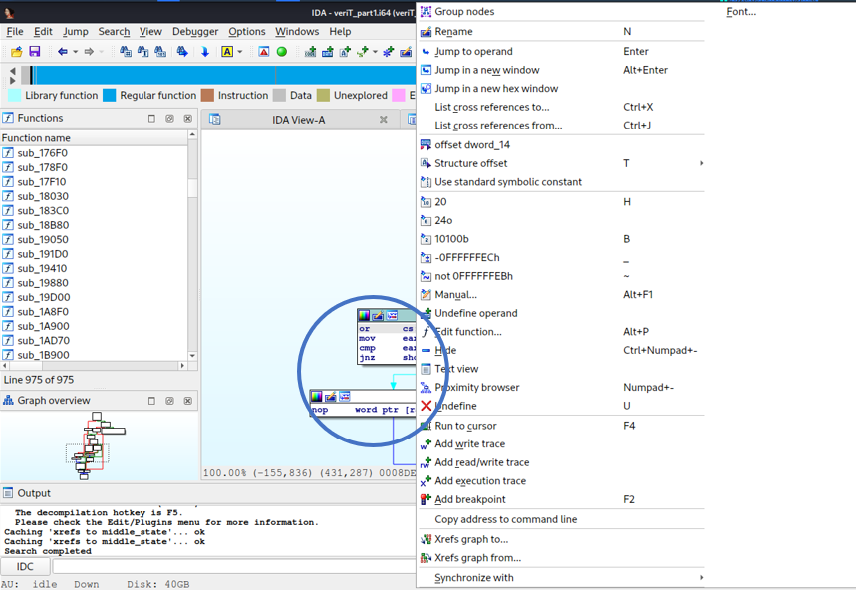}
\caption{\label{fig:false_negative_IDA_2}The IDA toolbar covers most of the screen and only a tiny portion of the function is visible. \changed{Occlusions like this make it difficult to identify symbols used for function annotations.}
 }   
\end{figure}

There are, however, some limitations. For instance, the tool may struggle to identify functions in screenshots where only a small portion of them is visible. In Figure~\ref{fig:false_negative_IDA}, the user has multiple windows open, revealing only a small section of an IDA function that, being zoomed in, does not contain sufficient relevant symbols. In Figure \ref{fig:false_negative_IDA_2}, the IDA toolbar covers most of the screen, making it difficult for the tool to identify any symbol used for function annotations. In both cases, the framework failed to detect any function. 

\begin{figure}[t]
\centering
\includegraphics[width=1\textwidth]{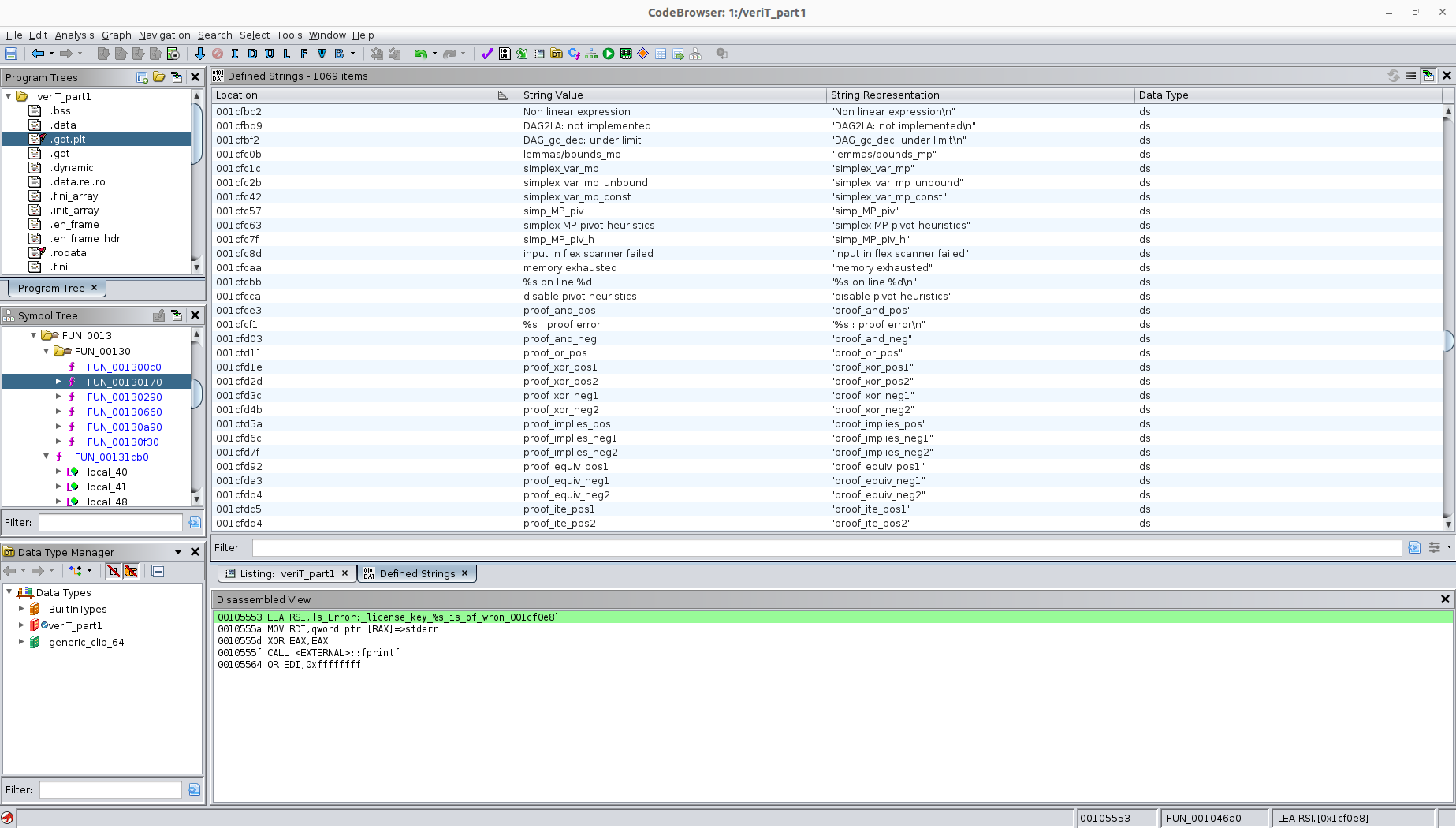}
\caption{\label{fig:incorrect_ghidra} \changed{In this screenshot, Ghidra's Defined Strings windows shows symbols that also exist as unique symbols in functions. As a result, the framework incorrectly classifies this screenshot as depicting a function.}}
\end{figure}

Mistakenly detecting a function when the screenshot has no function rarely occurs. One scenario, as shown in Figure~\ref{fig:incorrect_ghidra}, is that a user may open a Defined Strings window. Even though this screenshot does not contain any function, if such a window contains symbols that are also unique symbols in a function, the framework might incorrectly classify the screenshot as depicting a function. For example, printf format strings are typically unique to functions. However, a string table window will list all such strings, plus other strings, in a single window. 

Moreover, our framework currently does not support scenarios where more than one function appears in a single screenshot. Such scenarios are rather rare and typically occur only when the reverse engineer is scrolling or browsing through textual views (so-called Listing views) on the assembly code of a program, at which times parts of consecutive functions can be displayed together, rather than when they are conducting a detailed analysis of a function's code. We have not encountered this in any of the screenshots selected from our Datasets. Adding a feature to address this limitation would increase processing time, so we decided against it.

Occasional function annotation mistakes typically have minimal effects on the analysis procedures. In some instances such as in Figure~\ref{fig:false_negative_IDA_2}, the reverse engineer might not be actively examining the function's code at the time when the screenshot was taken. As discussed in Section~\ref{subsec:methodology}, for the purpose of this evaluation, we do not take these subjective observations into consideration. Additionally, as discussed in Section~\ref{subsub:function_matching}, when creating a timeline of RE activities, reAnalyst can consolidate nearby intervals by ignoring minor gaps, including gaps that result from function annotation inaccuracies. This means reAnalyst treats the time during the gap as being covered by the same function.

\subsection{Results for Basic Block Annotation Generation}
\label{subsec:basic_block_results}
  
\begin{table}[t]
\centering
\small
\begin{tabular}{ccccccc}
\textbf{Subject} &
  \textbf{Binary} &
  \textbf{Total} &
  \textbf{Correct} &
  \textbf{Incorrect} &
  \textbf{Undetected} & \textbf{Overall}\\
& & \textbf{\# Blocks} & \textbf{Annotations} & \textbf{Annotions} & \textbf{Blocks} & \textbf{Accuracy} \\
  \hline

1 & 1  & 87 & 85 & 1 & 1 & 96.7\% \\
2 & 1  & 78 & 76 & 0 & 2 & 96.5\% \\
3 & 2a & 179 & 176 & 1 & 2 & 97.3\%  \\
4 & 2b & 90 & 86 & 0 & 4 & 94.6\% \\
5 & 2c & 85 & 80 & 2 & 3 & 93.2\% \\
\hline
Total & & 519 & 503 & 4 & 12 & 95.9\%\\
\end{tabular}%
\caption{\label{table:basic_block_results} Breakdown of basic block annotation results for each experiment group. }
\end{table}

Table~\ref{table:basic_block_results} presents the accuracy obtained with our basic block annotation generation tool on the 20 IDA screenshots selected from 5 subjects in Dataset A. 
As we have excluded screenshots with incorrect function matches, when calculating the overall accuracy of the basic block annotation generation, we multiply the ratio of the number of correct identifications by the overall accuracy of the function annotation generation for Dataset A (from which these screenshots were taken) to get the numbers in the rightmost column.

Our basic block annotation generation tool exhibits good performance, achieving an overall accuracy rate of 95.9\%. This high accuracy rate signifies the framework's effectiveness in correctly matching basic blocks within the given screenshots, when applicable.

Still, our tool encounters challenges in specific cases. For instance, in Figure~\ref{fig:cropped_basic_blocks}, seven basic blocks are marked with letters A-G. Basic blocks D, E, and G are typical basic blocks fully visible on the screen that are hence easy to match. Basic blocks B and F are small, each with only one line, posing a challenge due to their similarity with other basic blocks in the program. Basic blocks C and A are partial blocks, with only a small portion visible. Although 4 out of the 7 basic blocks pose matching challenges, the framework correctly detected all of them except for basic block C, which has only a minimal portion visible. In such scenarios, our framework may fail to detect these partially visible basic blocks. Even when detection is successful, accurately matching them to the correct basic block may be difficult, because multiple basic blocks may share identical text and symbols, complicating the matching process.

\section{Future Work}\label{sec:futurework}

In our future work, we plan to refine our OCR-driven framework for better accuracy, especially in processing screenshots with complex layouts. An interesting research direction will consider scenarios in which reverse engineers use multiple analysis tools, and in which they keep all of them on display while their attention and interaction alternates between the tools. In such scenarios, we envision that multiple data sources, such as mouse positions, parts of the screens in which activity is spotted, and keystrokes, can be used to determine the focal point of the reverse engineer.

Whereas our past research focused mostly on extracting annotations for static analysis activities, we will shift our focus to dynamic analyses in the near future. Those techniques will at least include instrumentation and debugging.  

We will also consider combining information from multiple, consecutive screenshots to improve accuracy and obtain more information. For example, we may be able to detect patterns of actions that cannot easily be determined by analyzing screenshots in isolation, and we may be able to detect that a function is still being shown but zoomed out compared to the previous screenshot. In addition, our research focus will gradually shift towards higher-level analysis to extract more abstract annotation, such as the identification of mileposts in \changed{RE} strategies.

Integrating machine learning with our analysis framework could also provide deeper insights, including for developing predictive models. These could revolutionize our understanding of RE processes by forecasting likely next steps based on ongoing actions.

We also aim to develop comprehensive guidelines and best practices for efficiently organizing and conducting RE experiments, collecting data with our data collection framework, and analyzing collected data with reAnalyst. This will complement recently published recommendations on evaluation methodologies for software protections~\cite{desutter2024evaluation}. Our focus will be on streamlining the data collection, annotation generation, and analysis process, ensuring that it is as time-efficient and accurate as possible. By doing so, we hope to make the process more accessible and practical for researchers, enhancing their ability to analyze RE activities in depth and with effective and ethical practices and greater precision. This guideline will serve as a valuable resource for those utilizing our framework in various RE contexts.

\section{Availability}
\label{sec:availability}
Our contributions are open source and available online. The reAnalyst framework is available at \url{github.com/csl-ugent/reAnalyst}. The adapted version of the data collection framework can be accessed at \url{github.com/taylor239/UserMonitorServer}. We invite the community to explore, utilize, and improve these resources.

\section{Conclusions}
\label{sec:conclusions}
In this paper we present reAnalyst, a framework for the automatic generation of annotations
that identify reverse engineering activities. The input to our system is a stream of low-level
actions taken by a reverse engineer while solving a carefully designed challenge. In contrast to earlier work, our experimental setup allows for data to be collected on the subjects' machines, in the environment they are used to, using their preferred reverse engineering tool suites.

The generality of this experimental setup and the vast amount of raw data it collects cause problems for how the stream of data is ``coded'', i.e., how we map low-level actions to higher level annotations. Previous work has shown that manual analysis is infeasible.

In this paper we present, demonstrate, and evaluate a number of proof-of-concept tools
to automatically extract various annotations.

We provide evidence regarding the acceptability and practicality of our approach based on observations on how reAnalyst performs in various experiments, as well as through a survey conducted among reverse engineers.

We argue ---based on an extensive literature review--- that our approach can make future research into the practice of reverse engineering and the evaluation of software protections more accurate, more aligned with real-world reverse engineering practices, and more productive.

Compared to the most closely related work, our approach significantly improves over
RevEngE by Taylor, which only supported manual annotation of data streams, and over
ReMind by Mantovani et al., which only supported a custom
reverse engineering tool that offered severely limited functionality compared to real-world
RE tools.

\section{Funding}
We acknowledge support from The Research Foundation – Flanders (FWO) [Project nr.: 3G0E2318], from the Cybersecurity Research Program Flanders, and from the NSF under grants SATC/EDU-2029632 and SATC/TTP-1525820.

\bibliographystyle{spmpsci}
\bibliography{references}

\appendix
\section{Survey Questionnaire}\label{appendix:survey}

\changed{This appendix presents the complete set of survey questions referenced in Section~\ref{reversesurvey}. We used these survey questions to gather feedback from reverse engineers about the acceptability of our data collection framework.}

\subsection{Background and Experience}
\begin{enumerate}
    \item Have you previously used a virtual machine (VM) for reverse engineering software or related tasks?
    \item How many years of experience do you have in software reverse engineering?
    \item Which of the following best describes your primary area of work in software reverse engineering?
    \item If you were to participate in a reverse engineering experiment, are you employed by a company that would put restrictions on whether and how you can participate in such experiments?
\end{enumerate}

\subsection{Participation with Restrictions}
\begin{enumerate}
    \item In your role as an employee, which of the following best describes your employment setting?
    \item What is the size of your company/organization?
    \item Which of the following statements reflect your company's policies regarding your participation in reverse engineering experiments?
    \item (Optional) If some of the above restrictions would apply to you when participating in personal or professional capacity, can you describe them in more detail and explain why those restrictions are required by your company?
    \item Would your employer have concerns about installing third-party data collection software on company equipment, outside of a VM?
    \item Would the presence of the data collection framework cause you to alter your usual reverse engineering strategies or tools?
    \item Which of the following statements reflect your company policy regarding the public dissemination of results of reverse engineering experiments to which you participated in your professional capacity?
    \item What level of anonymization does your company require during data collection and analysis?
\end{enumerate}

\subsection{Participation without Restrictions}
\begin{enumerate}
    \item Which of the following best describes your occupation?
    \item Do you have any personal policies or ethical considerations that might restrict your participation in reverse engineering experiments or the use of certain tools during an experiment?
    \item Do you have concerns about installing third-party data collection software on your personal equipment, even within a VM?
    \item Would participating in this experiment conflict with any contractual obligations or agreements you have (e.g., with clients, partners)?
    \item Would the presence of the data collection framework cause you to alter your usual reverse engineering strategies or tools?
    \item Which of the following statements reflect your policy regarding your participation in reverse engineering experiments in your personal capacity?
    \item Which of the following statements would reflect your policy regarding the public dissemination of results of reverse engineering experiments to which you participated?
    \item What level of anonymization do you require during data collection and analysis?
\end{enumerate}

\subsection{Comfort and Concerns with Data Collection Framework}
\begin{enumerate}
    \item How comfortable would you feel using a data collection framework (that collects screenshots, keystrokes, etc.) during a reverse engineering experiment?
    \item What are your main concerns about using this data collection framework?
    \item Do you anticipate that using the data collection framework will affect your effectiveness or efficiency in reverse engineering tasks?
    \item If you work with malware samples, are you concerned about the framework affecting the integrity of malware analysis or interactions?
\end{enumerate}

\subsection{Conditions for Participation}
\begin{enumerate}
    \item What would make you more willing to participate in an experiment with data collection?
    \item What level of transparency do you require regarding data collection and usage?
    \item (Optional) Do you have any additional feedback or concerns about participating in a reverse engineering experiment using this data collection framework?
\end{enumerate}

\changed{
\section{Datasets Used in Evaluation}\label{appendix:datasets}
This appendix provides details on the datasets used in our evaluation as introduced in Section~\ref{sec:datasets}. We outline each dataset’s origin, the specific tasks assigned, and key differences between these datasets.

\paragraph{\textbf{Dataset A (Student experiment 2022, IDA Pro)}} We created four different binaries (Part 1 and Parts 2A, 2B, and 2C). All participants were asked to complete Part 1 and were then randomly assigned to one of the 3 binaries in Part 2. The free version of IDA Pro was their primary tool as they were trained to use it. For the challenge binaries, we embedded a license key checker within the open-source SMT solver verit, which comprises 52K lines of C source code. The license checker incorporates real and fake checks on subkeys, designed to execute at randomly chosen program points, at which point they access and update the global state of the license checker. 
We explained the general operation of the license checker to the students, i.e., the
automaton’s states and possible transitions, but we did not tell them what data structures
we used to store and encode this state. Their task was to extract the conditions that valid
keys need to meet by identifying the real checks during classroom sessions that ran for
approximately 3 hours. 
The Part 1 binary encoded the state in simple integers, the Part 2 binaries used hashmaps. The difference between the three Part 2 binaries was whether or not the injected license checker reused hashmap APIs and/or hashmap instances already in use in the main program. Our goal was to determine the impact of such so-called flexible software protection~\cite{jens2022flexible} on MATE attacks.

In total, 33 students participated and enabled the RevEngE data collection framework throughout the whole experiment.

\paragraph{\textbf{Dataset B (Student experiment 2023, Ghidra)}}  The binaries and assignments were similar to those in Dataset A, with only the encoding function of the automaton state being somewhat simpler to make the challenge easier. Moreover, in the 2023 edition of the course the students used Ghidra instead of IDA Pro. With this tool switch, the focus also switched from RE assembler CFGs to working with C code obtained through decompilation. Everything else remained the same, and participants were also asked to solve Part 1 and one of the three binaries in Part 2. A total of 29 students participated to the whole experiment.

\paragraph{\textbf{Dataset C (GREC 2021, Binary Ninja)}} Unlike the other two datasets, this dataset comes from a public challenge in which the participants were unconstrained; anyone was welcome to participate, at any time (during the two months of the experiment ran) of their choice, and using any RE tool of their choice. There were ten challenges of ten binaries and participants were eligible for monetary prizes for successfully solving the challenges with our data collection framework enabled. More information about the challenges and the binaries can be found online~\cite{grand2024challenge}. In the end, only two individuals successfully participated by solving at least one challenge. Of these, only one produced about 5 hours of data which we evaluated; the other successful participant used multiple devices, switched between challenges often, and had large intervals of inactive data where the subject had left their device idling. Due to these factors, it was difficult for us to accurately evaluate the performance of reAnalyst on that user. 
}
\end{document}